\documentclass[12pt,preprint]{aastex}
\newcommand{\PSbox}[3]{\mbox{\rule{0in}{#3}\includegraphics{#1}\hspace{#2}}}
\newcommand{\FigNum}[1]{\unitlength 1pt \begin{picture}(55,10)(-400,35)
                        \put(0,0){Figure #1}
                        \end{picture}}
\newcommand{\perval}[2]{{#1\mbox{$^{#2}$}}}

\newcommand{\persec}{\perval{\rm s}{-1}\/}
\newcommand{\percm}{\mbox{$\cm^{-2}$}}

\newcommand{\ppm}{\mbox{$\pm$}}
\newcommand{\cgsflux}{\erg\,\percm\,\persec}

\newcommand{\cgslum}{\erg~\persec}

\newcommand{\approxgt}{\mbox{$\gtrsim$}}

\newcommand{\nh}{\mbox{$N_{\rm H}$}}

\newcommand{\ee}[1]{\mbox{$10^{#1}$}}
\newcommand{\tee}[1]{\mbox{$\times 10^{#1}$}}

\newcommand{\keV}{\mbox{$\rm\,keV$}}

\newcommand{\cm}{\mbox{$\rm\,cm$}}

\newcommand{\erg}{\mbox{$\rm\,erg$}\/}

\newcommand{\mJy}{\mbox{$\rm\,mJy$}}

\newcommand{\Jy}{\mbox{$\rm\, Jy$}}

\newcommand{\rxte}{{\em RXTE\/}}

\newcommand{\kpc}{\mbox{$\rm\,kpc$}}

\newcommand\msec{\mbox{${\rm ms}$}}

\newcommand{\fxfr}{\mbox{$F_{X}/F_{\rm 1.4\,GHz}$}}
\newcommand{\fxlim}{\mbox{$F_{X, {\rm lim}}$}}

\def\rrat{RRAT~J1819$-$1458}

\received{}
\revised{}
\slugcomment{}

\begin{document}

\title{Constraints on RRAT Emission Mechanisms from {\em RXTE}/PCA
  Observations of \rrat}

\author{Robert E. Rutledge\altaffilmark{1} 
}

\altaffiltext{1}{Department of Physics, McGill University, 3600 rue
  University, Montreal, QC, H3A 2T8, Canada; rutledge@physics.mcgill.ca}

\begin{abstract}
We derive the second and most stringent limit to date of the
X-ray/radio flux ratio (\fxfr) for the radio bursts associated with
the recently identified source class, the Rotating Radio Transients
(RRATs).  We analyze 20.1 hr of \rxte/PCA observations of \rrat\ -- a
period during which 350\ppm23 RRAT radio bursts occurred, based on the
previously observed average radio burst rate. No X-ray bursts were
detected, implying an upper-limit on the X-ray flux for RRAT-bursts of
$\fxlim<$1.5\tee{-8} \cgsflux\ (2-10\,\keV), or a luminosity
$<2.3\tee{37} (d/3.6\kpc)^2$\cgslum.  The time-average burst flux is
$<$2\tee{-13} \cgsflux\ (0.5-8 keV) -- a factor of 10 below that of
the previously identified persistent X-ray counterpart. Thus, X-ray
bursts from the RRAT are energetically unimportant compared with the
persistent X-ray emission.  From the previously observed burst radio
flux, we derive an upper-limit $\fxfr\leq4.2\tee{-12}
\cgsflux\perval{\rm mJy}{-1}$ for the radio bursts from this RRAT, the
most stringent to date, due to the high radio flux of bursts from this
source.  The \fxfr\ ratio is a factor $\sim$80 larger than that of the
millisecond pulsar PSR~B1821-24; thus emission processes of
X-ray/radio efficiency comparable to MSP pulses cannot be ruled out.
However, if the RRAT burst emission mechanism is identical to the
\msec\ bursts of magnetars, then the \msec\ bursts of magnetars should
be easily detected with radio instrumentation; yet none have been
reported to date.
\end{abstract}

\keywords{stars: flare; stars: neutron; X-rays: stars; pulsars:
  individual (J1819$-$1458)}

\section{Introduction}

Recently, a new observational class of neutron stars has been proposed
(\citealt{mclaughlin06}, M06 hereafter), Comprised of eleven sources, the class is
defined by the phenomenon of fast (2-30 ms) radio bursts, which repeat
on timescales of 4\,min - 3\,hr.  Periodicities were claimed to be
found in the range of $P=$0.4-7\,s for ten of the eleven sources,
through period folding; this suggested association with rotating
neutron stars.  In three of the eleven sources, period derivatives
were claimed, with one suggesting a magnetic field strength of
5\tee{13} G -- a possible magnetar.  Because the $P-\dot{P}$ analysis
remains unpublished and undescribed, we regard association of the
radio-bursting population with neutron stars to be an unsupported
hypothesis; nonetheless, we adopt the hypothesis and nomenclature of
referring to the sources as Rotating RAdio Transients (RRATs), and
consider their association with neutron stars to be the leading
hypothesis.

One of these -- \rrat -- has a $7\arcsec\ \times32\arcsec$ error
ellipse (M06)\footnote{It is unclear why the positional uncertainty
  from this source is so much smaller than, for example, the
  $15\arcmin\times7\arcmin$ error region for RRAT J1911+00, obtained
  in the same study with the same instrumentation.}, which permitted discovery of an X-ray counterpart, CXOU
J181934.1$-$145804 \citep{reynolds06}, with absorption-corrected flux
of 2\tee{-12} \cgsflux\ (0.5-8 keV), or 3.1\tee{33} \cgslum\ at the
$d=3.6$\,\kpc\ dispersion-measure-indicated distance.  Two-hundred
twenty nine radio bursts at $\nu=$1.4 GHz, with average half-duration
$w_{50}$=3\,\msec\ were observed in 13 hours of radio integration, the
brightest with a peak flux of $S_\nu=3.6\,\Jy$.  If we take $w_{50}$
to be the light crossing time of the emission region, this implies a
source brightness temperature (e.g., \citealt{guedel02}):

\begin{equation}
T_b = 3\times10^{23} \left(\frac{S_\nu}{3.6\, {\rm Jy}}\right) 
\left( \frac{ 1.4\, {\rm GHz}}{\nu}\right)^2
\left( \frac{3\,{\rm ms}}{w_{\rm 50}} \right)^2
\left( \frac{d}{3.6\, {\rm kpc}} \right)^2 \, {\rm K }
\end{equation}

\noindent  This is significantly above the brightness temperature
limit for inverse Compton scattering ($T_B\sim\ee{12} K$)
which likely requires a coherent emission mechanism for this and
other RRATs, possibly accompanied by X-ray emission.

In this paper, we analyze \rxte/PCA data, and search for X-ray bursts
which may accompany the observed radio bursts, finding none.  In
\S~\ref{sec:obs}, we describe the observations and analysis; in
\S~\ref{sec:results}, we present the results of the X-ray burst
search; and in \S~\ref{sec:con}, we discuss the implications of these
results and conclude. 

\section{Observations and Analysis}
\label{sec:obs} 

We analyze observations taken in \rxte/PCA \citep{pca,pca2}, obtained
from the public archive (Table~\ref{tab:obs}).  For basic data
extraction, we used
FTOOLS\footnote{http://heasarc.gsfc.nasa.gov/ftools} \citep{ftools}
following standard \rxte\ recipes.  We used {\tt GoodXenon\_16s}-type
data, analyzing 20.1 hr of data.  We performed some data selections:
we excluded data from PCU1 (Proportional Counter Unit 1, numbered
0-4), and layers 2-3 from all PCUs, and required at least 2 active
PCUs.  Most observations subsequently had 3 or 4 active PCUs (0,2-4; or
0,2-3).  These selections eliminate the PCU high-voltage
breakdowns\footnote{HV breakdown in PCU3 and PCU4 is typically marked
by a burst of counts, usually in PHA channels $<$20, and limited to a
single wire within the PCU, on a $\leq2\msec$ timescale.  However,
breakdown in PCU1 is marked by low-PHA counts across {\em all} wires,
and is often terminated by a single high-energy (PHA 15-30) count; it
may be due to a distinct phenomenon.}  which occur in PCU1, PCU3 and
PCU4 \citep{pca2} in these and other data on which the programs were
tested.

We examined pulse-height-analyzer (PHA) channels only $\leq27$
corresponding to an approximate energy range of 1.5-10 keV, and
excluded propane-veto layer events.  

We searched for bursts on two timescales: $\tau=3.0$\,\msec\ (the
observed radio-burst timescale for \rrat) and 1/10 that value
(0.3\,\msec), to cover possible shorter X-ray duty cycles.

We measure the number of counts $M$ over a period of $\pm$1 sec about
each count in the data stream.  We measure the number $N$ of counts
which are observed within $\pm \tau/2$ of each count, and we calculate
the Poisson probability of measuring $\geq N$ counts for an expected
number of counts $\mu=M\times\tau/(2\, {\rm s})$ as:

$$
P_{\rm Poisson}(\mu, \geq N) = \sum_{i=N}^\infty \frac{\mu^N}{N!}\exp(-\mu)
$$

Summing over all 22 observations, we observed $N_{\rm phot}=3069964$
total counts.  An event for which $P_{\rm Poisson}\times N_{\rm
phot}<0.01$ would have 1\% chance of being produced due to random
fluctuation alone during the entirety of our analysis; we call such an
event a ``trigger'', and store all information about the $N+1$ counts
in that trigger for future examination.

\section{Results} 
\label{sec:results}
We found no triggers in the above data stream, on either timescale.  We
calculated the limits on burst rate as a function of average burst
flux over the 3\,ms and 0.3\,ms timescale (Fig.~\ref{fig:limits}).  At
the time of each count, we calculated the average (over $\tau$) number
of counts per PCU per sec, corresponding to a 99\% confidence
countrate fluctuation (that is, a Poisson fluctuation which would only
occur, on average, in a dataset with 100$\times$ as many counts as the
present dataset); this resulted in 99\% confidence upper-limits
between 422-1010 c/s/PCU on the $\tau=3\msec$ timescale).  During
these observations, \rrat\ lay 27.5\arcmin\ off-axis, at a collimator
efficiency of 0.45-0.5, for which we correct the countrates.  Layer-1
wires in the PCUs make up $\approx$90\% of the total active area below
10 keV, for which we correct the countrate (the remaining 10\% is in
layers 2-3, which we neglected).  Finally, we use an energy correction
factor of 1.0\tee{-8} \cgsflux (2-10\,\keV) per 1000 c/s/PCU (assuming
a photon power-law slope $\alpha$=1, and neglecting \nh\footnote{using
WebPIMMS; http://heasarc.gsfc.nasa.gov/Tools/w3pimms.html} which we
assume as the default spectrum throughout), where here, the countrates
are for layers 1-3.  The maximum burst rate is the 95\% confidence
upper-limit for the burst rate, $2./T(>F_{\rm lim})$, where $T(>F_{\rm
lim})$ is the total integrated observing time sensitive to bursts from
\rrat\ with flux $>F_{\rm lim}$; the numerator (2) comes from the
number of bursts for which one detects 0 bursts $\leq$5\% of the time.

In general, flux sensitivity did not change much during observations;
thus, we detected no bursts on a $\tau=3\,\msec$ ($\tau=0.3\,\msec$)
timescale at fluxes above $\approxgt1.5\tee{-8} \cgsflux$ ($8\tee{-8}
\cgsflux$), with 99\% confidence, with burst rates as low as
0.1\,\perval{hr}{-1}.  At lower fluxes, the observations rapidly
become insensitive.

\section{Discussion and Conclusions}
\label{sec:con}

\rrat\ is the most prolific radio-burster of all the reported RRATs --
229 radio bursts detected in 13\,hr of radio observations.  On
average, we would expect 350\ppm23 such radio bursts to have occurred
during these observations, assuming the radio burst rates do not
change with time.  None of these had a corresponding X-ray flux
$\geq$1.5\tee{-8} \cgsflux (2-10 keV).  We expect 2, on average, to
have been as radio bright as the brightest previously observed (3.6
Jy), which implies that $\geq$1 such bursts were emitted (95\%
confidence).  This, in combination with our derived burst flux limit,
results in an X-ray to radio flux ratio limit for bursts from \rrat\ of
approximately:

$$
\fxfr \leq 4.2\tee{-12}\, \cgsflux\,{\rm mJy}^{-1}
$$

\noindent This limit is a factor of 14 lower than that derived for
RRAT~J1911+00 \cite{hoffman06}, due to the factor of 14 greater radio
flux observed from the brightest radio burst from \rrat\
(M06).  It may be instructive to compare this ratio
with that from other transient coherent emission processes.

The limit corresponds to \fxfr$\leq\ee{14.7}$ Hz, comparable to the average
ratio  observed from stellar coronal activity
\citep{guedel93}; however the timescales for coronal flares (for
example) are typically $\sim10^3\,\sec$, much longer than the \msec\
flares of RRATs. 

Compared to MSPs, this upper-limit to \fxfr\ is a factor 84 above the
observed ratio for (for example) the millisecond pulsar PSR~B1821-24
($\fxfr = 5\tee{-14} \cgsflux \perval{\rm mJy}{-1}$;
\citealt{hoffman06}).   Thus, this limit does not exclude an emission
mechanism for RRAT bursts with comparable X-ray to radio efficiency to
MSPs emission mechanism. 

The claim of measured $P-\dot{P}$ values (M06) which imply magnetic
field strengths approaching those of magnetars, makes comparison
between the RRAT-bursts and those from magnetars a natural one, with
the expectation that the emission mechanisms may be similar.  The
absence of X-ray/radio burst flux limits for the \msec\ X-ray bursts
of magnetars
\citep{gogus99,gogus01,gavriil02,kaspi03,gavriil04,woods05} precludes
direct comparison with the emission mechanism powering those sources.
However, a radio observation of such would be valuable to compare with
the derived \fxfr\ limits for RRATs.  Taking 1E~2259+586 as an example
\citep{gavriil04}, with median duration of $\sim$100 \msec, and peak
fluences of \ee{-9} \erg\perval{\rm\,cm}{-2}, magnetar bursts would
produce coincident radio bursts of $\geq2500\mJy$, if the magnetar
bursts were due to a similar emission mechanism as RRATs. This radio
flux is comparable to the brightest RRAT radio bursts yet observed
(M06), and are therefore readily detectable, yet no such bursts have
been reported.  However, the burst rate of the \msec\ X-ray bursts of
magnetars is not constant in time; whether this is true for the radio
bursts of RRATs has not be described in the reported observations
(M06).  On the other hand, a non-detection of radio bursts from
magnetars to flux limits below this value would require a different
\fxfr\ ratio for the \msec\ bursts from magnetars, and may therefore
imply a distinct emission mechanism.

A burst flux limit $F_{\rm lim}<$1.5\tee{-8} \cgsflux (2-10 keV),
averaged over 3\,\msec, combined with an average burst rate of 228/13
hr (M06), results in time-average burst flux of
$<$2.2\tee{-13} \cgsflux\ (2-10 keV), or, extrapolating to the energy
band for the X-ray counterpart \citep{reynolds06} $<$2.1\tee{-13}
\cgsflux (0.5-8 keV).  The time-average burst X-ray flux is a factor
of 10 less than the persistent X-ray counterpart in this pass-band
\cite{reynolds06}.   Thus, X-ray bursts related to the radio
bursts of this RRAT are energetically unimportant compared to the
persistent X-ray flux.  In addition, the time-average burst X-ray
efficiency $\eta=L_{X}(0.5-8\,\keV)/\dot{E}$, compared with the
reported dipolar emission $\dot{E}=24.94(5)\tee{31} \cgslum$
(M06)  is $\eta\leq7\tee{-3} (d/3.6\,kpc)^2$.

It is unlikely that great improvement will be made in the X-ray flux
limit presented here using existing instrumentation.  An improved flux
limit cannot be obtained with {\em Chandra}: 2 counts in the HRC-S
detected within 2\,\msec\ corresponds to 3\tee{-8} \cgsflux
(extrapolated to 2-10 keV).  On-source \rxte/PCA observations could
improve on the present flux sensitivity by a factor of $\times2$.

\acknowledgements

RER acknowledges support from NSERC through the Discovery Grant
program.  RER acknowledges useful conversations with Marjorie
Gonzalez, Margaret Livingstone, Vicky Kaspi and Andrew Cumming.

\clearpage

\begin{figure}[htb]
\caption{ \label{fig:limits} 
Upper-limits on burst rate as a function of flux, averaged on a 3\,\msec\
(solid line) and 0.3\,\msec\ timescales, from \rrat\ (see
\S~\ref{sec:results} for discussion). 
}
\end{figure}

\clearpage
\pagestyle{empty}
\begin{figure}[htb]
\PSbox{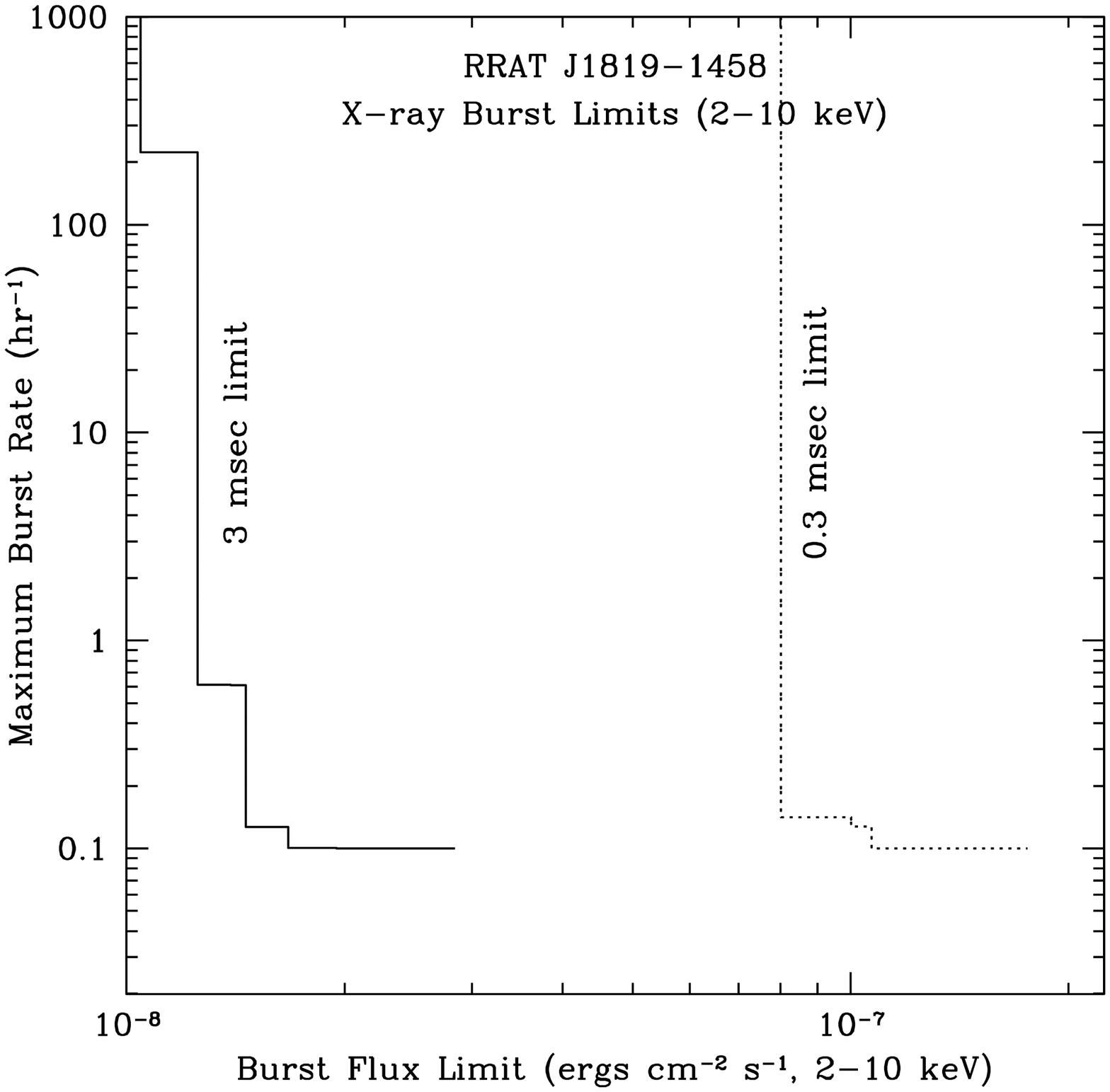 hoffset=-80 voffset=-80}{14.7cm}{21.5cm}
\FigNum{\ref{fig:limits}}
\end{figure}

\clearpage

\clearpage

\begin{deluxetable}{lccrcr}
\tablecaption{\rxte\ Observations of \rrat\ \label{tab:obs} }
\tablehead{
\colhead{Start Time (TT)} &
\colhead{End Time (TT)} &
\colhead{} &
\colhead{} &
\colhead{} &
\colhead{ $\langle I \rangle$ (c/s)$^a$} \\
\colhead{(dd/mm/yy hh:mm)} &
\colhead{(dd/mm/yy hh:mm) } &
\colhead{(ksec)} &
\colhead{\# PCUs$^b$} &
\colhead{$N_{\rm phot}$} &
\colhead{(1.5-10 keV)}\\
}
\startdata
  06/06/98   12:50 &   06/06/98   13:42 &  3.16 &   4 &  155048  &   49.07   \\ 
  06/06/98   14:25 &   06/06/98   15:25 &  3.63 &   4 &  183683  &   50.57   \\ 
  06/06/98   16:02 &   06/06/98   17:01 &  3.53 &   4 &  182342  &   51.57   \\ 
  06/06/98   17:42 &   06/06/98   18:37 &  3.31 &   4 &  159157  &   48.05   \\ 
  08/06/98   09:39 &   08/06/98   10:25 &  2.73 &   4 &  124533  &   45.52   \\ 
  08/06/98   11:15 &   08/06/98   12:01 &  2.73 &   4 &  119602  &   43.71   \\ 
  08/06/98   12:51 &   08/06/98   13:43 &  3.13 &   4 &  143406  &   45.73   \\ 
  08/06/98   14:27 &   08/06/98   15:26 &  3.55 &   4 &  167412  &   47.13   \\ 
  08/06/98   16:03 &   08/06/98   17:03 &  3.61 &   4 &  172435  &   47.69   \\ 
  08/06/98   17:43 &   08/06/98   18:40 &  3.39 &   4 &  161687  &   47.67   \\ 
  10/06/98   12:54 &   10/06/98   13:45 &  3.02 &   4 &  129359  &   42.78   \\ 
  10/06/98   14:28 &   10/06/98   15:27 &  3.56 &   4 &  158155  &   44.33   \\ 
  10/06/98   16:04 &   10/06/98   17:05 &  3.64 &   4 &  158619  &   43.48   \\ 
  10/06/98   19:21 &   10/06/98   20:17 &  3.34 &   2 &  140593  &   42.06   \\ 
  10/06/98   20:58 &   10/06/98   21:53 &  3.26 &   4 &  131917  &   40.42   \\ 
  10/06/98   22:34 &   10/06/98   23:14 &  2.43 &   4 &   97358  &   40.03   \\ 
  12/06/98   11:18 &   12/06/98   12:02 &  2.67 &   3 &   98495  &   36.88   \\ 
  12/06/98   12:53 &   12/06/98   13:46 &  3.21 &   4 &  134874  &   41.94   \\ 
  12/06/98   14:29 &   12/06/98   15:28 &  3.55 &   4 &  120253  &   33.86   \\ 
  12/06/98   16:05 &   12/06/98   17:05 &  3.61 &   3 &  113601  &   31.42   \\ 
  12/06/98   17:41 &   12/06/98   18:41 &  3.63 &   3 &  111473  &   30.69   \\ 
  12/06/98   19:19 &   12/06/98   20:17 &  3.50 &   3 &  105962  &   30.24   \\ 
\enddata
\tablenotetext{a}{Average countrate during observation for PHA
  channels $\leq$27.}
\tablenotetext{b}{Number of active PCUs at observation start time.
  PCUs turn on and off during the course of the observations.}
\end{deluxetable}


\begin{thebibliography}{}

\bibitem[\protect\astroncite{{Blackburn}}{1995}]{ftools}
{Blackburn}, J.~K., 1995,
\newblock in R.~A. {Shaw}, H.~E. {Payne}, \& J.~J.~E. {Hayes} (eds.), {\em ASP
  Conf. Ser. 77: Astronomical Data Analysis Software and Systems IV}, p. 367

\bibitem[\protect\astroncite{{Gavriil} {\rm et~al.\/}}{2002}]{gavriil02}
{Gavriil}, F.~P., {Kaspi}, V.~M., \& {Woods}, P.~M., 2002,
\newblock {\em \nat} { 419}, 142

\bibitem[\protect\astroncite{{Gavriil} {\rm et~al.\/}}{2004}]{gavriil04}
{Gavriil}, F.~P., {Kaspi}, V.~M., \& {Woods}, P.~M., 2004,
\newblock {\em \apj} { 607}, 959

\bibitem[\protect\astroncite{{G{\"o}{\u g}{\"u}{\c s} } {\rm
  et~al.\/}}{1999}]{gogus99}
{G{\"o}{\u g}{\"u}{\c s} }, E., {Woods}, P.~M., {Kouveliotou}, C., {van
  Paradijs}, J., {Briggs}, M.~S., {Duncan}, R.~C., \& {Thompson}, C., 1999,
\newblock {\em \apjl} { 526}, L93

\bibitem[\protect\astroncite{{G{\"o}{\u g}{\"u}{\c s}} {\rm
  et~al.\/}}{2001}]{gogus01}
{G{\"o}{\u g}{\"u}{\c s}}, E., {Kouveliotou}, C., {Woods}, P.~M., {Thompson},
  C., {Duncan}, R.~C., \& {Briggs}, M.~S., 2001,
\newblock {\em \apj} { 558}, 228

\bibitem[\protect\astroncite{{G{\"u}del}}{2002}]{guedel02}
{G{\"u}del}, M., 2002,
\newblock {\em \araa} { 40}, 217

\bibitem[\protect\astroncite{{Guedel} \& {Benz}}{1993}]{guedel93}
{Guedel}, M. \& {Benz}, A.~O., 1993,
\newblock {\em \apjl} { 405}, L63

\bibitem[\protect\astroncite{{Hoffman} {\rm et~al.\/}}{2006}]{hoffman06}
{Hoffman}, K., {Rutledge}, R.~E., {Fox}, D.~B., {Gal-Yam}, A., \& {Cenko},
  S.~B., 2006,
\newblock {\em \apj}, submitted, astro-ph/0609092

\bibitem[\protect\astroncite{{Jahoda} {\rm et~al.\/}}{2006}]{pca2}
{Jahoda}, K., {Markwardt}, C.~B., {Radeva}, Y., {Rots}, A.~H., {Stark}, M.~J.,
  {Swank}, J.~H., {Strohmayer}, T.~E., \& {Zhang}, W., 2006,
\newblock {\em \apjs} { 163}, 401

\bibitem[\protect\astroncite{{Kaspi} {\rm et~al.\/}}{2003}]{kaspi03}
{Kaspi}, V.~M., {Gavriil}, F.~P., {Woods}, P.~M., {Jensen}, J.~B., {Roberts},
  M.~S.~E., \& {Chakrabarty}, D., 2003,
\newblock {\em \apjl} { 588}, L93

\bibitem[\protect\astroncite{{McLaughlin} {\rm et~al.\/}}{2006}]{mclaughlin06}
{McLaughlin}, M.~A., {Lyne}, A.~G., {Lorimer}, D.~R., {Kramer}, M., {Faulkner},
  A.~J., {Manchester}, R.~N., {Cordes}, J.~M., {Camilo}, F., {Possenti}, A.,
  {Stairs}, I.~H., {Hobbs}, G., {D'Amico}, N., {Burgay}, M., \& {O'Brien},
  J.~T., 2006,
\newblock {\em \nat} { 439}, 817

\bibitem[\protect\astroncite{{Reynolds} {\rm et~al.\/}}{2006}]{reynolds06}
{Reynolds}, S.~P., {Borkowski}, K.~J., {Gaensler}, B.~M., {Rea}, N.,
  {McLaughlin}, M., {Possenti}, A., {Israel}, G., {Burgay}, M., {Camilo}, F.,
  {Chatterjee}, S., {Kramer}, M., {Lyne}, A., \& {Stairs}, I., 2006,
\newblock {\em \apjl} { 639}, L71

\bibitem[\protect\astroncite{{Swank} {\rm et~al.\/}}{1996}]{pca}
{Swank}, J.~H., {Jahoda}, K., {Zhang}, W., \& {Giles}, A.~B., 1996,
\newblock in M.~A. Alpar, U. Kiziloglu, \& J. Van~Paradijs (eds.), {\em The
  Lives of the Neutron Stars}, NATO ASI Ser C., 450), Boston: Kluwer

\bibitem[\protect\astroncite{{Woods} {\rm et~al.\/}}{2005}]{woods05}
{Woods}, P.~M., {Kouveliotou}, C., {Gavriil}, F.~P., {Kaspi}, V.~M., {Roberts},
  M.~S.~E., {Ibrahim}, A., {Markwardt}, C.~B., {Swank}, J.~H., \& {Finger},
  M.~H., 2005,
\newblock {\em \apj} { 629}, 985

\end{thebibliography}
\end{document}